\documentclass[hyper]{PoS}

\usepackage{textgreek}
\usepackage{graphicx}
\DeclareGraphicsExtensions{.pdf,.png,.jpg}
\usepackage{subeqn}
\usepackage{cancel}

\newtoks\hreftoks
\makeatletter
\@ifundefined{texorpdfstring}{}{}
\@ifundefined{href}{\newcommand\href[2]{#2}}
\@ifundefined{doi}{\newcommand\doi[2]{\hreftoks{#1}%
         \href{http://dx.doi.org/\the\hreftoks}{#2}}}{}
\@ifundefined{arXiv}{\newcommand\arXiv[1]{%
         \href{http://arXiv.org/abs/#1}{arXiv:#1}}}{}
\makeatother

\title{Neutron Electric Dipole Moment from Beyond the Standard Model}

\ShortTitle{Neutron Electric Dipole Moment from Beyond the Standard Model}

\author{\speaker{Tanmoy Bhattacharya}\\
        Los Alamos National Laboratory and the Santa Fe Institute\\
        E-mail: \email{tanmoy@lanl.gov}}

\author{Vincenzo Cirigliano and Rajan Gupta\\
        Los Alamos National Laboratory}

\abstract{We discuss the phenomenology of neutron Electric Dipole Moment
         from the Standard Model and beyond, and identify the matrix
         elements most necessary to connect the current and forthcoming 
         experiments with phenomenology.  We then describe lattice
         techniques for calculating these matrix elements.}

\FullConference{The 30th International Symposium on Lattice Field Theory\\
		 June 24--29,  2012\\
		 Cairns, Australia}

\makeatletter
\@ifundefined{Hy@AtBeginDocumentHook}{}
  {\Hy@AtBeginDocumentHook\global\let\Hy@AtBeginDocumentHook\relax}
\makeatother

\renewcommand\pdfsetcolor[1]{}

\begin{document}

\section{Introduction}
The observed universe has \(6.1^{+0.3}_{-0.2}\times 10^{-10}\)
baryons for every black body photon~\cite{WMAP}, whereas in a baryon
symmetric universe, we expect no more that about \(10^{-20}\) baryons
for every photon~\cite{KolbTurner}.  It is difficult to include such a
large excess of baryons as an initial condition in an inflationary
cosmological scenario~\cite{NoInflation}.  The way out of the impasse
lies in generating the baryon excess dynamically during the evolution
of the universe.  Sakharov wrote down a set of three necessary
conditions for such a process to be possible: baryon number violation,
CP and T violation, and out of equilibrium evolution of the
universe~\cite{Sakharov}. Efforts at generating an observable baryon
excess when these conditions are not satisfied have not been
promising~\cite{CPreviews}. Since every Lorentz invariant field
theory action needs to be symmetric under the product CPT~\cite{CPT},
we use CP-violation and T-violation interchangeably and ignore the
possibility of explicit Lorentz violations or spontaneous
breaking of CPT.\looseness-1

CP is violated in the standard model (SM) of particle physics by a
phase in the Cabibo-Kobayashi-Maskawa quark mixing
matrix~\cite{CKMphase}, and possibly by a similar phase in the
leptonic sector if the neutrinos are not
massless~\cite{Neutrinophase}. The physical effects of these phases
are supressed by the smallness of the fermion masses. Baryon number is
also violated in the SM by sphaeleron effects in weak
interactions~\cite{Sphaeleron}, though the difference of baryon and
lepton numbers is strictly conserved unless the neutrinos have a
Majorana mass.  At the temperatures above the electroweak transition,
where the sphaeleron rates are high, baryon and antibaryons,
therefore, equilibriate.  Since, however, the electroweak phase
transition is weakly first order, the universe never goes out of
equilibrium enough to generate the observed baryon
density through SM processes~\cite{CPreviews}.

In principle, the SM has an additional source of CP violation arising
from the effect of QCD instantons.  The presence of these finite
action non-perturbative configurations in a non-abelian theory often
leads to inequivalent quantum theories defined over various
`$\Theta$'-vaccua~\cite{Theta}. Because of asymptotic freedom, all
non-perturbative configurations including instantons are strongly
suppressed at high temperatures where baryon number violating
processes occur. Because of this, CP violation due to such vaccuum
effects do not lead to appreciable baryon number production.

This analysis points to the need to look for CP violation from beyond
the standard model (BSM).  A promising experimental approach is to
measure the static electric dipole moments of elementary particles,
which are necessarily proportional to their spin. Since under
time-reversal spin reverses sign but the electric dipole moment does
not, a non-zero measurement would imply CP violation.  In this work,
we concentrate on the electric dipole moment of the neutron (nEDM).

\section{Operators}
In the SM, CP violation arises from (i) the CKM phase in the charged
current weak interactions, and (ii) the \(\Theta\)-term multiplying
the topological charge density operator in the strong interaction
sector. In this section, we discuss the phenomenology of these and the
leading BSM operators.

\subsection{CKM phase}
\label{sec:CKM}
The CKM matrix describing quark-mixing under charged current
interactions is arbitrary up to quark field redefinitions that leave
the rest of the Lagrangian unchanged. In the SM, this can be used to
rotate away any phase in the CKM matrix unless there are at least
three non-degenerate generations of up and down type quarks, with
non-zero values for the sines and cosines of the three mixing
angles~\cite{Gilman}.  Because of this, the CKM contribution to the
quark electric dipole moment (qEDM) are suppressed by the quark mass
differences and mixing angles, and is only \(O(10^{-34})\) e~cm
because of further partial cancellations between three-loop
diagrams~\cite{Czarnecki}.  The contribution to nEDM from weak diquark
interactions within the neutron have been estimated to be a 100 times
larger than this~\cite{Dar}, but is still far below experimental
sensitivity of \(O(10^{-28})\) e~cm.

\subsection{Topological Charge}
\label{sec:Q}
Even though the QCD \(\Theta\)-term does not give rise to appreciable
baryon number violation, it does contribute to the electric dipole
moment of baryons.  The \(U(1)_A\) axial chiral anomaly, however,
allows one to simultaneously change the \(\Theta\)-term and redefine
the fermion fields by a chiral phase, without changing the physics. In
the SM, such a rotation can remove all physical effects of a similar
term in the weak \(SU(2)\) sector that involves only the left chiral
sector of the theory. For a vector theory like QCD, however, all the
quarks obtain masses through the Higgs mechanism, and a rotation to
remove \(\Theta\) introduces phases into this mass matrix.  The
fermion phases are then conventionally chosen such that the masses are
positive and real.  The value of \(\Theta\) under this choice of
phases is called \(\bar\Theta\), and is phenomenologically known to be
close to zero rather than \(\pi\)~\cite{Crewther}.

For small \(\bar\Theta\), its contribution to nEDM has been estimated
to be \(5.2\times10^{-16} \bar\Theta\) e~cm in chiral peturbation
theory~\cite{Crewther}.  Since the current experimental limit is
\(2.9\times10^{-26}\) e~cm~\cite{Experimental}, this implies
\(|\bar\Theta| \lesssim 10^{-10}\). One can explain such an
unnaturally small value by the `Peccei-Quinn' (PQ) construction of
elevating \(\Theta\) to a dynamical field: in the SM, the minimum of
the potential is at PQ field \(\bar\Theta = 0\).\looseness-1

\subsection{Beyond the standard model}
\label{sec:BSM}

\begin{figure}[t]
\begin{center}
\includegraphics[width=\textwidth]{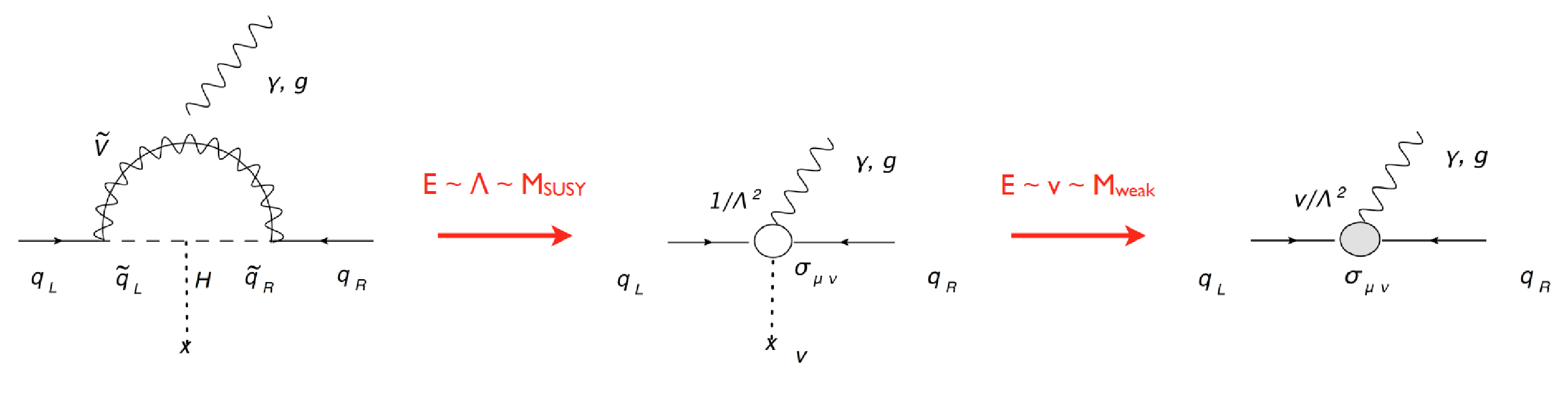}
\end{center}
\caption{One loop diagram involving squarks and neutralinos in a SUSY
  model that can give rise to a electric dipole moment or a
  chromo-electric dipole moment to a quark. The external wavy line
  denotes a generic gauge boson that can attach at various places in
  the diagram. \(\Lambda\) represents the heavy SUSY scale where the
  loop effectively becomes pointlike, whereas \(v\) is the electroweak
  scale where the Higgs vacuum expectation value breaks the
  electroweak symmetry.}
\label{fig:BSMqEDM}
\end{figure}

BSM theories can be parameterized by the effective low energy
operators they introduce.  Electric dipole moments of quarks are
particularly interesting operators since, as discussed in
Sec.~\ref{sec:CKM}, these operators arise only at three-loops in
SM. In BSM, they can arise at one-loop (Fig.~\ref{fig:BSMqEDM}) and
produce nEDMs of about \(2.9 \times 10^{-26}\)
e~cm~\cite{Experimental}.  Also, since they violate chirality, a phase
rotation to make the masses real typically mixes the electric and
magnetic dipole operators.  CP violating electric dipole moments
coupling to electromagnetic, weak, and strong gauge bosons are,
therefore, generic unless the dipole moment sector is precisely
`aligned' with the mass sector.

Though formally of mass dimension 5, such chirality violating
operators can appear only due to the breaking of electroweak
symmetry. One, therefore, expects them to be suppressed by the ratio
of the weak scale to the scale of BSM physics.  Unsupressed operators
appear at dimension 6 and typically involve four fermions or three
powers of the gauge field strength tensor.

\section{Model Estimates}
As explained in Sec.~\ref{sec:BSM}, after choosing the field basis to
make the mass terms real and positive, the major contributions to the
nEDM come from the quark electric and chromoelectric dipole moment
operators in addition to topological charge effects induced by
\(\Theta\):
\begin{eqnarray}
{\cal S}^{\cancel{\hbox{CP}}} &=&  \Red{-i\, \bar\Theta \frac{g^2}{16\pi^2} {
                         \ G^{\mu\nu a} \tilde G_{\mu\nu a} }}
                                      \nonumber \\
        &\Green{+}& \Green{i\, e \, d_u^\gamma {\bar L}
                                        \sigma_{\mu\nu}\gamma_5
                                        F^{\mu\nu} \frac{\tilde
                                          H}{\tilde v} U 
         +  i\, e\, d_d^\gamma {\bar L}
         \sigma_{\mu\nu} \gamma_5 F^{\mu\nu} \frac Hv D} \nonumber \\
        &\Orange{+}& \Orange{i\, g \, d_u^G {\bar L}
             \sigma_{\mu\nu}\gamma_5 \lambda^a G^{\mu\nu\, a} 
               \frac{\tilde H}{\tilde v} U 
         +  i\, g \,d_d^G {\bar L}
         \sigma_{\mu\nu} \gamma_5\lambda^a G^{\mu\nu\, a} \frac Hv D}
         + \cdots \ ,
\end{eqnarray}
where \(e\) and \(g\) are the electromagnetic and strong couplings,
\(F\) and \(G\) represent the electromagnetic and
gluonic field strengths, \(L\) represents the left handed quark
fields, \(U\) and \(D\) represent the right handed up-type and
down-type quarks, \(H\) and \(\tilde H\) represent possibly distinct
Higgs doublets whose vacuum expectation values, \(v\) and \(\tilde v\)
(assumed real), break the electroweak symmetry, and
\(d_{u,d}^{\gamma,G}\) represent the electric and chromoelectric
dipole moments of the up and down quarks.

Model estimates of the effect of these terms show
\begin{subequations}
\begin{eqnarray}
d_n &\approx& \frac{8 \pi^2}{M_n^3}\left[
   \Red{-\frac{2 m_*}{3} \frac{\partial \langle \bar q \sigma q\rangle_F}{\partial F}
       \left(\bar\Theta \Orange{{} + g \frac{\langle \bar q G \sigma q \rangle}{2\langle \bar q q\rangle}
       \sum \frac {d_q^G}{m_q}}\right)}\right.\label{eq:PQ}\\
   &&\qquad\Green{{}+\frac{\langle\bar qq\rangle}3\left(4\, d_d^\gamma - d_u^\gamma\right)} 
     \left.\Orange{{}+g \frac{\langle \bar q G \sigma q \rangle}{6\langle \bar q q\rangle}
      \left(
      4\,d_d^G\,\frac{\partial \langle \bar d \sigma d\rangle_F}{\partial F}
      - d_u^G\,\frac{\partial \langle \bar u \sigma u\rangle_F}{\partial F}
      \right)}
    \right]\\[\baselineskip]
    &\approx& \Green{\left(\frac43 d_d^\gamma - \frac 13d_u^\gamma\right)} 
              \Orange{{}-\frac{2e\langle \bar qq\rangle}{M_nf_\pi^2}
                      \left(\frac23 d_d^G + \frac13 d_u^G\right)}\,,
\end{eqnarray}
\end{subequations}
where \(M_n\) is the nucleon mass, \(m_*\) is the reduced light quark
mass, \(f_\pi\) is the pion decay constant, \(\langle\bar qq\rangle\)
and \(\langle\bar q G\sigma q\rangle\) represent the quark condensate
and the mixed quark-gluon condensate, respectively, and \(\langle\bar
d\sigma d\rangle_F\), \(\langle\bar u\sigma u\rangle_F\) and \(\langle
\bar q\sigma q\rangle_F\) represent the up, down and average tensor
light quark condensates induced by an external uniform
electromagnetic field \(F\)~\cite{Pospelov}.  In the simplified
expression one assumes that the term in Eq.~(\ref{eq:PQ}) vanishes by
the PQ mechanism. Numerically, one finds:\looseness-1
\begin{subequations}
\begin{eqnarray}
d_n(\bar\Theta) &\approx&
             (1 \pm 0.5) \frac{|\langle\bar qq\rangle|}{(225\rm MeV)^3}
                   \Red{\bar\Theta\,(2.5 \times 10^{-16}\, \hbox{e~cm})}\\
d_n(d_q^{\gamma,G}) &\approx& \Orange{-d_n\left(\bar\Theta
  %=\Theta_{\rm ind}
  \approx \sum \frac{d_q^G/m_q(\rm MeV)}{(3.1 \times 10^{-17}\rm \hbox{e~cm})}
                           \frac{|m_0^2|}{(0.8\rm GeV)^2}
  \right) + {}} \label{eq:ind} \\ &&
                (1 \pm 0.5) \frac{|\langle\bar qq\rangle|}{(225\rm MeV)^3}
                  \left[\Orange{ 1.1\,(d_d^G + 0.5\,d_u^G)\,e + {}}\right.
               \left.\Green{
                  1.4\,(d_d^\gamma-0.25\,d_u^\gamma)}\right]\,,
\end{eqnarray}
where 
% \begin{equation}
%  % 0.8 GeV^2/MeV = 640000 MeV = (3.1E-4 fm)^-1 
%   \Orange{
%    \Theta_{\rm ind} \approx (3.1 \times 10^{-17}\rm cm)^{-1}
%                           \sum \frac{d_q^G}{m_q(\rm MeV)}
%                            \frac{|m_0^2|}{(0.8\rm GeV)^2}
%                            }
% \end{equation}
\(m_0^2 \equiv \langle\bar q G\sigma q\rangle/\langle\bar qq\rangle\)
and Eq.~(\ref{eq:ind}) is a contribution that is cancelled in PQ
theory because of the non-zero \(\bar\Theta\) at the minimum of the
BSM potential~\cite{Pospelov}. The quark dipole moments are often
of the order of
\begin{equation}
% 1 MeV/16pi^2 1TeV^2 = 6.3E-14 MeV^-1 = 1.3E-12 fm
  \kappa_q \equiv \frac{m_q}{16\pi^2 M_\Lambda^2} = 
  1.3\times10^{-25} \hbox{e~cm} \frac{m_q}{1\rm MeV} 
  \left(\frac{1 TeV}{M_\Lambda}\right)^2\,,
\end{equation}
where \(M_\Lambda\) is the scale of new physics. 

Rough estimates of the other dimension 6 operators, the Weinberg
operator \((w/3) f_{abc} G^{\mu\nu,a}\allowbreak G^{\rho,b}_\mu\tilde
G_{\rho\nu}^c\) and the four-quark operators \(i C_{ij} (\bar
q^a_iq^a_i)(\bar q^b_j\gamma_5q^b_j)\), are also similar~\cite{Pospelov}:
\begin{eqnarray}
% 22 MeV/1TeV^2 = 2.2E-11 MeV = 4.4E-9 fm
|d_n(w)| &\approx& (4.4 \times 10^{-22}\,\hbox{e~cm}) 
                        \;\left.\frac{w(\mu)}
                               {(1\rm TeV)^{-2}}\right|_{\mu=1\rm GeV}\\
% 2.6E-3 GeV^2 / 4.2 GeV/1TeV^2 = 6.2E-13 MeV^-1 - 1.2E-11 fm 
|d_n(C)| &\approx& (1.2 \times 10^{-24}\,\hbox{e~cm})
               \;\left.\frac{C_{bd}(\mu) + C_{db}(\mu)}
                            {(1 \rm TeV)^{-2}}\right|_{\mu=m_b}\,.
\end{eqnarray}
\end{subequations}

\section{Lattice Methods}

In this section we discuss the calculation of the contribution of
three CP violating operators---the topological charge, the electric
dipole moment of the quark, and the chromo-electric dipole moment of
the quark---to nEDM.  Since nEDM changes the energy of the neutron in
an external electric field, there are two general ways of calculating
it---as the energy difference between two spin states of a neutron in
an external electric field,
  % \begin{equation}
  %   d_n = \frac12 \left.\left(M_{n\downarrow} -
  %       M_{n\uparrow}\right)\right|_{E=E\uparrow}\,,
  % \end{equation}
  % where \(d_n\) is the nEDM, \(E\uparrow\) is a constant electric
  % field in the `up' direction, and \(M_{n\uparrow}\) and
  % \(M_{n\downarrow}\) are the energies of a zero momentum neutron with
  % spin parallel and anti-parallel to the electric field,
  % respectively. 
or by relating it to the CP-violating form factor \(F_3\):\looseness-1
%  \begin{subequations}
  \begin{equation}
    \left.\langle n \mid J^{\rm EM}_\mu \mid n \rangle\right|_{\cancel{\hbox{CP}}} = \frac{F_3(q^2)}{2
      M_n} \bar n q_\nu \sigma^{\mu\nu} \gamma_5 n\qquad
    d_n = \lim_{q^2\to0} \frac {F_3(q^2)}{2 M_n}\,,\label{eq:F3}
    \end{equation}
%    \end{subequations}
    where \(J^{\rm EM}_\mu\) is the electromagnetic current, \(\bar
    n\) and \(n\) are appropriate spinors for the neutron, and \(q\) is
    the momentum transfer in the 
    process. We will only consider the second method here.

In either of these cases, we need to calculate lattice correlation
functions in the presence of a CP violating operator.  This is
technically difficult because the CP violating operator is complex,
so, in practice one needs to expand the action for small values of the
CP-violation parameter. 
% In particular, this is accomplished by the
% identity 
% \begin{equation}
% \langle C^{\cancel{\hbox{CP}}}(x,y,\ldots)\rangle =
% -\langle C^{\cancel{\hbox{CP}}}(x,y,\ldots) {\cal L}^{\cancel{\hbox{CP}}}
% (p_\mu=0)\rangle_{\hbox{CP}}\,,
% \end{equation}
% where \(C^{\cancel{\hbox{CP}}}(x,y,\ldots)\) is a CP-violating correlation function
% depending on arbitrary space time points \(x,y,\ldots\), \({\cal
%   L}^{\cancel{\hbox{CP}}}(p_\mu=0)\) is the CP violating part of th
% Lagrangian evaluated at zero momentum and \(\langle\ldots\rangle_{\hbox{%
% CP}}\) means the correlation calculated under the corresponding
% CP-conserving action.

\subsection{Topological Charge}

For the \(\Theta\)-term, the CP violating part of the action is
% \({\cal L}^{\cancel{\hbox{CP}}}(p_\mu=0) =
\(\Theta\int d^4x G_{\mu\nu}\tilde G^{\mu\nu} = \Theta Q\),
where \(Q\) is the topological charge.  Hence, we need to evaluate
\begin{eqnarray}
  \left.\langle n \mid J^{\rm EM}_\mu \mid n \rangle\right|_{\cancel{\hbox{CP}}} &=& \Theta \left\langle n
    \left| \left( \frac23 \bar u \gamma_\mu u - \frac13 \bar
        d\gamma_\mu d\right) Q \right| n \right\rangle\nonumber\\
  &=& \frac\Theta2 \left\langle n \left| \bar q \gamma_\mu q Q \right| n
  \right\rangle + \frac\Theta6 \left\langle n \left| \bar q \gamma_\mu
  \tau_3 q Q \right| n \right\rangle\,,
\end{eqnarray}
where the formulae are written for a mass-degenerate two-flavour
theory with a doublet (\(q\)) consisting of up (\(u\)) and down
(\(d\)) quark fields and all matrix elements on the rhs are calculated
in a CP conserving background lattices generated with \(\Theta=0\).
Since the topological charge is independent of the quark propagators,
this reduces to weighted sums of three-point functions in each
topological sector. 

\begin{figure}[t]
\begin{tabular}{p{0.45\textwidth}p{0.45\textwidth}}
\begin{center}
\includegraphics[width=0.2\textwidth]{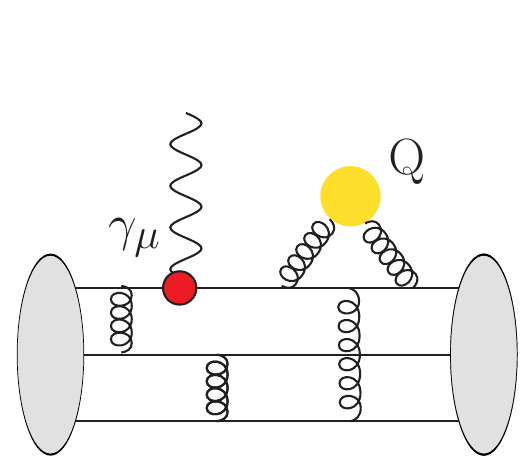}
\includegraphics[width=0.2\textwidth]{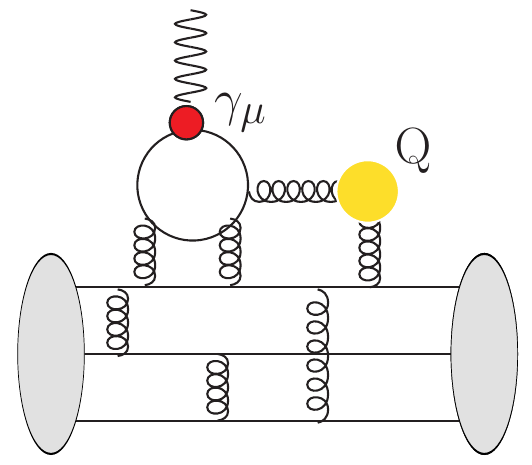}
\end{center}&
\begin{center}
\includegraphics[width=0.2\textwidth]{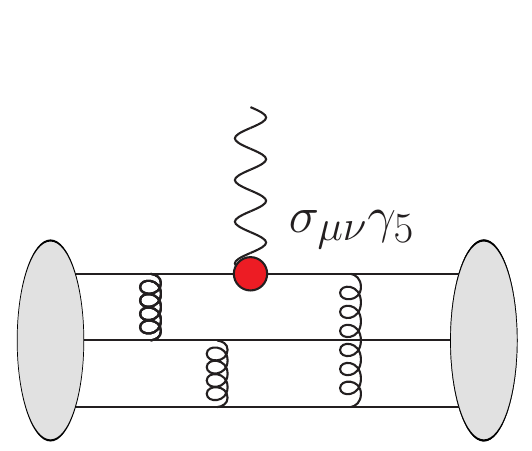}
\includegraphics[width=0.2\textwidth]{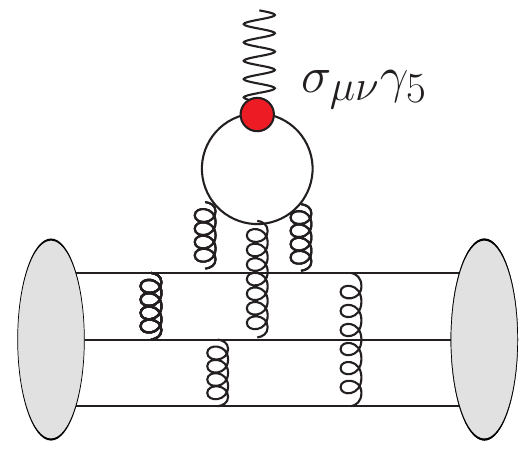}
\end{center}\\
\caption{The lattice matrix elements that are needed to calculate the
  contribution of the \(\Theta\)-term to nEDM. 
  % The left
  % `connected' diagram only contributes to the isoscalar part, whereas
  % the right `disconnected' diagram contributes to both the isoscalar
  % and isovector terms.
  The circle labeled `Q' represents that each diagram is
  to be weighted by the topological charge of the configuration.}
\label{fig:Theta}
&%
\caption{The lattice matrix elements that are needed to calculate the
  contribution of the quark electic dipole moment to nEDM.
  % The left
  % `connected' diagram only contributes to the isoscalar part, whereas
  % the right `disconnected' diagram contributes to both the isoscalar
  % and isovector terms.
}
\label{fig:qEDM}
\end{tabular}
\end{figure}

To isolate the CP-violating form factor, these matrix elements have to
be calculated at non-zero momentum, and then the limit of zero
momentum taken. The two lattice diagrams contributing to this process
are shown in Figure~\ref{fig:Theta}.

\subsection{Quark Electric Dipole Moment}

When the up and down quarks have non-zero electric dipole moments they
directly give extra CP violating contributions to the electromagentic
current.  As a result, the nEDM is given by
\begin{eqnarray}
  \left.\langle n \mid J^{\rm EM}_\mu \mid n \rangle\right|_{\cancel{\hbox{CP}}} &=& \left\langle n
    \left| \left( d_u^\gamma \bar u \sigma_{\mu\nu} u + d_d^\gamma \bar
        d\sigma_{\mu\nu} d\right) q^\nu \right| n \right\rangle\nonumber\\
  &=& q^\nu \frac{d_u^\gamma+d_d^\gamma}2 \left\langle n \left| \bar q \sigma_{\mu\nu} q \right| n
  \right\rangle + q^\nu \frac{d_u^\gamma-d_d^\gamma}2 \left\langle n
  \left| \bar q \sigma_{\mu\nu} \tau_3 q
      \right| n \right\rangle\,.\label{eq:qEDM}
\end{eqnarray}
Even though Eq.~(\ref{eq:qEDM}) suggests that the calculation needs
injection of non-zero momentum (\(q^\nu\)) at the operator, note that
the form factor is also multiplied by the same factor in
Eq.~(\ref{eq:F3}). As a result, this calculation can be performed
directly at zero momentum.

The effect of the quark electric dipole moments, therefore, turn out
to be related to the iso-scalar and iso-vector tensor charges of the
nucleon that have been studied in other contexts in the
past~\cite{gT}. The two lattice diagrams are shown in
Figure~\ref{fig:qEDM}.

% \begin{figure}[tbh]
% \end{figure}

\subsection{Quark Chromoelectric Dipole Moment}

The contribution of the Chromoelectric dipole moments are more
difficult to evaluate, since they na\"\i vely need us to evaluate a four
point function, the technology for which has not yet been tested on
the lattice.  We can, however, evaluate it using the Feynman-Hellmann
theorem:
\begin{eqnarray}
  \left.\langle n \mid J^{\rm EM}_\mu \mid n \rangle\right|_{\cancel{\hbox{CP}}} &=& \left\langle n
    \left| J^{\rm EM}_\mu \left( d_u^G \bar u \sigma_{\nu\kappa} u + d_d^G \bar
       d\sigma_{\nu\kappa} d\right) \tilde G^{\nu\kappa} \right| n \right\rangle\nonumber\\
    &=& \left.\frac\partial{\partial A_\mu(E)}\right|_{E=0}
         \left.\left\langle n \left| \left( d_u^G \bar u \sigma_{\nu\kappa} u + d_d^G \bar
          d\sigma_{\nu\kappa} d\right) \tilde G^{\nu\kappa} \right| n \right\rangle\right|_E\,,
\end{eqnarray}
where the subscript \(E\) refers to matrix elements calculated in the
presence of an external electric field \(E\), and \(A_\mu(E)\) refers
to the corresponding vector potential. Since the background electric
field breaks translational invariance, no momentum needs to be
explicitly introduced at the operator.\looseness-1

\section{Conclusions}

In this article, we have studied the calculation of the nEDM due to
the leading operator contributing to CP-violation and arising from
beyond the standard model physics.  Prior work~\cite{LatticeTheta} had
concentrated on the effects and renormalization of the so-called
\(\Theta\)-term.  Here, we show that the effect of the quark electric
dipole moment is related to the tensor charge of the neutron, which
has also been previously studied~\cite{gT}. We also describe the
technique for calculating the effect of the quark chromo-electric
moment, and postpone the discussion of the renormalization of this
term.  Some of the four-fermion operators recently discussed in the
literature~\cite{fourfermi} are also calculable with similar effort,
but we do not discuss them here.

This work was supported by DOE grant nos.~DE-KA-1401020 and
DE-AC52-06NA25396.  The lattice calculations described here
are being performed in collaboration with Saul D. Cohen, Anosh Joseph,
and Huey-Wen Lin.

\end{document}